\documentstyle[12pt]{article}
\bigskip\bigskip
\begin{document}

\bigskip
\bigskip
\bigskip

\centerline{\Large A Generalization of Grover's Algorithm}
\bigskip
\bigskip
\centerline{  Luigi  Accardi  }
\centerline{  Ruben Sabbadini}  \bigskip

\centerline{   Centro Vito Volterra}

\centerline{    Universit\`a degli Studi di Roma ``Tor Vergata''}
\centerline{    Via Orazio Raimondo, 00173 Roma, Italia }
\centerline{    accardi@volterra.mat.uniroma2.it, http://volterra.mat.uniroma2.it }

\bigskip\bigskip\bigskip
\centerline{  ABSTRACT}
\bigskip
We investigate the necessary and sufficient conditions in order that a 
unitary operator can amplify a pre-assigned component relative to a 
particular basis of a generic vector at the expence of the other components.  
This leads to a general method which allows, given a vector and one of its 
components we want to amplify, to choose the optimal unitary operator which 
realizes that goal. Grover's quantum algorithm is shown to be a particular 
case of our general method.

However the general structure of the unitary we find is remarkably similar 
to that of Grover's one: a sign flip of one component combined with a 
reflection with respect to a vector. In Grover's case this vector is fixed; 
in our case it depends on a parameter and this allows optimization.
 
\bigskip

\section{Unitary operators which increase the probability of the $|0>$ 
component of a pre-assigned vector}

\bigskip
\bigskip
Let $|i> (i=0, \dots , N-1)$ be an orthonormal basis of $R^N$.
The mathematical core of Grover's algorithm is the construction of a 
unitary operator U which increases the probability of one of the
components of a given unit vector, in the given basis, at the expence of 
the remaining ones. The necessity of such an amplification of probabilities 
arises in several problems of quantum computation. For example in the 
Ohya-Masuda [4] quantum SAT algorithm such a problem arises. In a recent 
interesting paper Ohya and Volovich have proposed a new method of 
amplification, 
based on non linear chaotic dynamics [14]. In the present paper we begin 
to study the following problem: is it possible to extend Grover's algorithm 
so that  it becomes applicable to a more general class of initial vectors, 
for example those wich arise in the Ohya-Masuda algorithm?
A preliminary step to solve this problem is to determine the most general 
unitary operator which performs the same task  of Grover's operator. This 
is done in Theorem (1.1) below. The result is rather surprising: we find that, 
up to the choice of four $\pm 1$  (phases), there exists exactly one class of
 such unitary operators, labeled by an arbitrary parameter in the interval 
$[0, 1]$. Moreover these unitaries can be written in a form similar 
to Grover's one, i.e. a reflection with respect to a given unit vector possibly preceeded by a sign flip of one component combined with, 
where the unit vector in question depends on this parameter 
in $[0, 1]$. The free parameter in our problem allows to solve a new 
problem, which could not be formulated within the framework of Grover's 
explicit construction, namely the \emph{optimization  problem} with 
respect to the given parameter. We prove that, even in the case of 
Grover's original algorithm, this additional freedom allows to speed 
up considerably the amplification procedure. In a forthcoming paper [15] we 
plan to apply the present method to the Ohya-Masuda algorithm. Since an 
operator $U$ is unitary if and only if it leaves unaltered the scalar 
products of vectors with real components in a given basis, we shall 
restrict our attention to unitary operators with real coefficients in a 
given basis (as the Grover's ones). This restrictes the problem to $  R^N$.
\bigskip\bigskip

\noindent{ THEOREM 1}\ \  Given the linear functionals:
\begin{equation}
\eta: a=(a_i)\in R^N\mapsto\eta (a)=\sum_{i=0}^{N-1}\eta_ia_i\label{gr1}
\end{equation}
\begin{equation}
c:a=(a_i)\in R^N\mapsto c(a)=\sum_{i=0}^{N-1}\gamma_ia_i\label{gr2}
\end{equation}
with $\gamma_i$ and $\eta_i$ real and $\varepsilon_1$, $\varepsilon_2 \in
\left\{\pm 1\right\}$, necessary and sufficient condition for the 
operator U, defined by:
\begin{equation}
U\sum a_i|i>=\varepsilon_1(a_0+\eta (a))|0>+\varepsilon_2\sum_{i\not=0}
\left( a_i+c(a)\right)|i> \label{gr3}
\end{equation}
to be unitary is that there exist a real number $\beta_0$ such that:
\begin{equation}
|\beta_0|\le 1\label{gr4}
\end{equation}
\begin{equation}\gamma_0\ \ =\varepsilon_5{\sqrt{1-\beta_0^2}\over \sqrt{N-1}}
\label{gr5}
\end{equation}
\begin{equation}
\gamma_i\ \ =-{{1+\varepsilon_3\beta_0}\over{N-1}}\qquad i\not=0\label{gr6}
\end{equation}
\begin{equation}
\eta_0\ \ =-1+\varepsilon_4\beta_0\label{gr7}
\end{equation}
\begin{equation}
\eta_i\ \ =\varepsilon_3\gamma_0\qquad i\not=0\label{gr8}
\end{equation}
where $\varepsilon_3$, $\varepsilon_4$, $\varepsilon_5$ are arbitrarily chosen in the
set $\{\pm1\}$. 
\bigskip

\noindent{PROOF}\ \ In finite dimension unitarity is equivalent to
isometry. Therefore U is unitary if and only if, for every $|a>=
\sum_{i=0}^{N-1}a_i |i>$ the following isometricy condition is satisfied:
$$\sum a_i^2=(a_0+\eta)^2+\sum_{i\not=0}(a_i+ c)^2=a_0^2+
\eta^2+2a_0\eta+\sum_{i\not=0}a_i^2+(N-1)c^2+2c\sum_{i\not=0}a_i$$
where we write $\eta$, $c$ for $\eta(a)$, $c(a)$. This condition can be 
written in the form:
\begin{equation}
\eta^2+2a_0\eta+(N-1)c^2+2c\sum_{i\not=0}a_i=0\label{gr9}
\end{equation}

With the notation:
\begin{equation}
\gamma(a)=\gamma:=(N-1)c^2+2c\sum_{i\not=0}a_i\label{gr10}
\end{equation}

Equation (\ref{gr9}) is equivalent to:
\begin{equation}
\eta^2+2a_0\eta+\gamma=0\label{gr11}
\end{equation}
and its possible solutions are:
\begin{equation}
\eta(a)=\eta=-a_0+\varepsilon_4\sqrt{a_0^2-\gamma(a)}\label{gr12}
\end{equation}

Given (\ref{gr12}) the  funtional $\eta (a)$ will be linear if and only 
if $\forall a_0, \dots , a_N$:

\begin{equation}
a_0^2-\gamma (a)=\left(\sum_j\beta_ja_j\right)^2\label{gr13}
\end{equation}
for some real numbers $\beta_j$ indipendent of $a$.

Since the functional $c(a)$ is linear and given by (\ref{gr2}), because 
of (\ref{gr12}) and (\ref{gr13}), condition (\ref{gr9}) becomes:

$$-a_0^2+(N-1)\left(\sum_j\gamma_ja_j\right)^2+\left(\sum_j\beta_ja_j\right)^2
+ 2\sum_j\gamma_ja_j\sum_{i\not=0}a_i=$$
$$-a_0^2+2\sum_j\gamma_ja_j\sum_{i\not=0}a_i+\sum_{i,j}\left[ (N-1)\gamma_i
\gamma_j+ \beta_i\beta_j\right]a_ia_j=0$$
or equivalently:
$$a_0^2\left[ (N-1)\gamma_0^2+\beta_0^2-1\right]+\sum_{i,j\not= 0}
\left[ 2\gamma_j+(N-1)
\gamma_i\gamma_j+\beta_i\beta_j\right]a_ia_j+$$
\begin{equation}
+2\sum_{i\not= 0}\left[ \gamma_0+(N-1)
\gamma_0\gamma_i+\beta_0\beta_i\right]a_0a_i =0\label{gr14}
\end{equation}

The identity (\ref{gr14}) holds $\forall a_0, \dots , a_N$, if and only if:

\begin{equation}
(N-1)\gamma_0^2+\beta_0^2-1=0\label{gr15}
\end{equation}
\begin{equation}
2\gamma_j+(N-1)\gamma_i\gamma_j+\beta_i\beta_j=0\ \quad\forall i,j\not=0\quad 
i\not=j\label{gr16}
\end{equation}
\begin{equation}
2\gamma_i+(N-1)\gamma_i^2+\beta_i^2=0\quad\ \forall i\not=0\label{gr17}
\end{equation}
\begin{equation}
\gamma_0+(N-1)\gamma_0\gamma_i+\beta_0\beta_i =0\quad\ \forall i\not=0
\label{gr18}
\end{equation}

Equation (\ref{gr15}) and the reality condition on $\eta$ imply that  
(\ref{gr4}) and (\ref{gr5}) hold. From (\ref{gr18}) we deduce that, 
for $i\not=0$:
\begin{equation}
\gamma_i=-{{\gamma_0+\beta_0\beta_i}\over{\gamma_0(N-1)}} \label{gr19}
\end{equation}
and, replacing this into (\ref{gr17}), we find:
$$-{{2(\gamma_0+\beta_0\beta_i)}\over{\gamma_0(N-1)}}+{{(\gamma_0+
\beta_0\beta_i)^2}\over{\gamma_0^2(N-1)}}+\beta_i^2=0$$
or:
$$\left[ \left(N-1\right)\gamma_0^2+\beta_0^2\right]\beta_i^2=\gamma_0^2$$
which, because of (\ref{gr15}), is equivalent to:
\begin{equation}
\beta_i=\varepsilon_3\gamma_0=\varepsilon_3\varepsilon_5{\sqrt{1-\beta_0^2}\over \sqrt{N-1}}
\label{gr20}
\end{equation}
with $\varepsilon_3=\pm 1$. Replacing (\ref{gr20}) into (\ref{gr19}) we arrive
to (\ref{gr6})

Replacing (\ref{gr25}),$\dots$, (\ref{gr28})  into (\ref{gr1}) and 
(\ref{gr2}), we conclude that a necessary condition for the linearity of 
U is that $\eta$ and $c$ must have the form:
\begin{equation}
\eta (a)=(-1+\varepsilon_4\beta_0)a_0+\varepsilon_4\varepsilon_3\gamma_0\sum_{k\not=0}a_k=
(-1+\varepsilon_4\beta_0)a_0+\varepsilon_4\varepsilon_3\varepsilon_5{\sqrt{1-\beta_0^2}\over \sqrt{N-1}}
\sum_{k\not=0}a_k\label{gr22}
\end{equation}
\begin{equation}
c(a)=\gamma_0a_0-{{1+\varepsilon_3\beta_0}\over{N-1}}\sum_{k\not=0}a_k=
\varepsilon_5{\sqrt{1-\beta_0^2}\over \sqrt{N-1}}a_0-{{1+\varepsilon_3\beta_0}\over{N-1}}
\sum_{k\not=0}a_k\label{gr23}
\end{equation}

Conversely, if conditions (\ref{gr4}), \dots, (\ref{gr8}) are satisfied, 
then also (\ref{gr14}), which is equivalent to (\ref{gr9}), is satisfied 
and therefore U is isometric, hence unitary. This can also be seen
by a direct computation (see appendix A).
\bigskip

\noindent{REMARK}\ \ Because of (\ref{gr4}) there exists a $\theta
\in \left[ 0, 2\pi\right)$  such that $\beta_0$ has the form:
\begin{equation}
\beta_0=\varepsilon_3cos\ \theta\label{gr24}
\end{equation}
and therefore, from (\ref{gr5}):
\begin{equation}
\sqrt{N-1}\gamma_0=\varepsilon_5\sqrt{1-\beta_0^2}=sin\ \theta\label{gr25}
\end{equation}
i.e. the parameters $\beta_0$ and $\gamma_0$ live onto an ellipse in the 
$(\beta_0,\ \gamma_0)$-plane. With these notations one has:

\begin{equation}
\eta (a)=\left(-1+\varepsilon_3\varepsilon_4 cos\ \theta\right)a_0+\varepsilon_3\varepsilon_4{sin\
\theta\over\sqrt{N-1}}\sum_{k\not=0}a_k\label{gr26}
\end{equation}

\begin{equation}
c(a)={{sin\ \theta}\over\sqrt{N-1}}a_0-{{1+ cos\ \theta}\over {N-1}}
\sum_{k\not=0}a_k\label{gr27}
\end{equation}

\bigskip
\bigskip

\noindent{REMARK}\ \ \ The case $\gamma=0$ leads to $\eta=0$ or
 $\eta=-2a_0$; in both cases we have:
$$U\sum a_i|i>=\pm \varepsilon_1a_0\ |0> + \varepsilon_2\sum_{i\not=0}(a_i+c)\ |i>$$
The operators
$U(\gamma\equiv0(a))$ are in this class, however they play a
significant role in Grover's algorithm
because they may be used to  change the sign of a component leaving the 
others $u$ inaltered (\emph{flip}).

If we are interested in unitaries which modify the component $a_0$ of $a$, 
we must look for solutions with $\gamma\not= 0$.
\bigskip

\noindent{ COROLLARY 2}\ \ If in (\ref{gr22}) and (\ref{gr23}) we
choose:

$$\varepsilon_1\varepsilon_4=\varepsilon_3=\varepsilon_5=1$$
$$\varepsilon_2=-1$$
$$\beta_0={{N-2}\over N}$$
$$\gamma_0={2\over N}$$
then the corresponding operator U is Grover's unitary (see section 4).
\bigskip

\noindent{PROOF}
It is known that Grover's unitary is characterized by (see section 4):
\begin{equation}
a_0\mapsto {{N-2}\over N}a_0+{2\over N}\sum_{k\not=0}a_k=:\varepsilon_1\left
[a_0+\eta (a)\right]\label{gr28}
\end{equation}
\begin{equation}
a_i\mapsto -a_i+{2\over N}\left(-a_0+\sum_{k\not=0}a_k\right)=:\varepsilon_2
\left[a_i+c(a)\right]\label{gr29}
\end{equation}

On the other hand, from equations (\ref{gr26}) and (\ref{gr27}) we have:
\begin{equation}
\varepsilon_1\left[ a_0+\eta(a)\right]
=\varepsilon_1\varepsilon_4\left(\beta_0a_0+\varepsilon_3\gamma_0\sum_{k\not=0}a_k\right)
\label{gr30}
\end{equation}
\begin{equation}
\varepsilon_2\left[ a_i+c(a)\right] =\varepsilon_2\left(a_i+\gamma_0a_0-{{1+\varepsilon_3
\beta_0}\over{N-1}}\sum_{k\not=0}a_k\right)\label{gr31}
\end{equation}
with $\gamma_0$ given by (\ref{gr5}). Comparing this with (\ref{gr28}) 
and (\ref{gr29}) we see that the condition for equality is:
$$\varepsilon_1\varepsilon_4\beta_0={{N-2}\over N}$$
Now let us choose $\varepsilon_1\varepsilon_4=1$ and $\beta_0={{N-2}\over N}$ then:
$$\gamma_0=\varepsilon_5\sqrt{1-{{(N-2)^2}\over{N^2}}\over{N-1}}=\varepsilon_5{2\over N}$$
that leads to $\varepsilon_5=1$. Therefore, if $\varepsilon_2=-1$, the coefficient of
the third term in (\ref{gr31}) becomes:
$$-\varepsilon_2{{1+\varepsilon_3\beta_0}\over {N-1}}={{1+\varepsilon_3{{N-2}\over N}}\over
{N-1}}= {{N+\varepsilon_3 N-2\varepsilon_3}\over {N(N-1)}}$$
that gives the correct parameter ${2\over N}$ if and only if $\varepsilon_3=1$.
\bigskip

\section{Canonical form and reflections} 
\bigskip\bigskip

\noindent{ THEOREM 1}\ \ Any unitary operator $U(\beta_0,\varepsilon)$
with real coefficients in the basis $(|i>)$ and satisfying the conditions 
of  Theorem (1.1), can be written in the form:
$$U(\beta_0,\varepsilon):=\varepsilon_1\varepsilon_4|0> \left( \beta_0\ <0|+\varepsilon_3\varepsilon_5
{\sqrt{1-\beta_0^2}\over \sqrt{N-1}}  \sum_{k\not=0}<k|\right)+$$
\begin{equation}
+\varepsilon_2\sum_{i\not=0}|i>\ \left( <i|+\varepsilon_5{\sqrt{1-\beta_0^2}\over
\sqrt{N-1}}<0|-{{1+\varepsilon_3\beta_0}
\over {N-1}}\  \sum_{k\not=0}<k|\right)\label{gr32}
\end{equation}

Moreover a unit vector $u\in R^N$ such that:
\begin{equation}
U(\beta_0,\varepsilon)=\varepsilon_2\left(1-2|u><u|\right)\label{gr33}
\end{equation}
exists if and only if $\varepsilon$ is such that
\begin{equation}
\varepsilon_2=\varepsilon_1\varepsilon_4\varepsilon_3\label{gr33a}
\end{equation}
In this case $|u>$ has the form:
\begin{equation}
|u>={1\over\sqrt{2}}\left|-\varepsilon_5\sqrt{1-\varepsilon_3\beta_0},\
\sqrt{{1+\varepsilon_3\beta_0}\over{N-1}},\ \dots ,\ \
\sqrt{{1+\varepsilon_3\beta_0}\over{N-1}}\right\rangle\label{gr34}
\end{equation}

\noindent{REMARK}\ \ Notice that unitary operator (\ref{gr33}) simply
realizes the reflection of the $|u>$-component of any vector with respect to 
the $|u>$-axis.

\noindent{PROOF}
The identity (\ref{gr32}) follows immediately from (\ref{gr2}), (\ref{gr22}) 
and (\ref{gr23}).

The operator $U(\beta_0,\varepsilon)$ of the equation (\ref{gr32}) can be
rapresented in the following way:
$$U(\beta_0,\varepsilon):=\varepsilon_21-\left\{|0>\left[\left(\varepsilon_21-
\varepsilon_1\varepsilon_4\beta_0\right) <0|-\varepsilon_1\varepsilon_4\varepsilon_3\varepsilon_5{\sqrt{1-\beta_0^2}
\over \sqrt{N-1}}\sum_{k\not=0}<k|\right]\right.+$$
\begin{equation}
+\left.\varepsilon_2\sum_{i\not=0}|i>\ \left(-\varepsilon_5{\sqrt{1-\beta_0^2}\over
\sqrt{N-1}}<0|+{{1+\varepsilon_3\beta_0}
\over {N-1}}\  \sum_{k\not=0}<k|\right)\right\}\label{gr34}
\end{equation}

Now an easy calculation shows that, given a vector $|u>$ of the form 
(\ref{gr34}), the right end side of (\ref{gr33}) is equal to:
$$\varepsilon_21-\left\{ |0>\left[\varepsilon_2\left( 1-\varepsilon_3 \beta_0\right)
 <0|-\varepsilon_2\varepsilon_5{\sqrt{1-\beta_0^2}\over \sqrt{N-1}}\sum_{k\not=0}
<k|\right]\right.+$$
\begin{equation}
+\left.\varepsilon_2\sum_{i\not=0}|i>\ \left((-\varepsilon_5{\sqrt{1-\beta_0^2}\over
\sqrt{N-1}}<0| +{{1+\varepsilon_3\beta_0}\over {N-1}}\  \sum_{k\not=0}<k|\right)
\right\}\label{gr35}
\end{equation}

For $\beta_0\not= 0,1$ (\ref{gr35}) and (\ref{gr34}) are equal if and only 
if (\ref{gr33a}) holds.
and the last operator is a projector if and only if:
\begin{equation}
\varepsilon_2=\varepsilon_1\varepsilon_4\varepsilon_3\label{gr36}
\end{equation}
because the off-diagonal terms must be equal. From this the thesis follows 
observing that multiplying (\ref{gr36}) for $\varepsilon_2\varepsilon_3$ we obtain:
\begin{equation}
\varepsilon_3=\varepsilon_1\varepsilon_4\varepsilon_2\label{gr37}
\end{equation}

\bigskip

\noindent{ COROLLARY 2}\ \ Grover's operator is the product of a operator
of the form (\ref{gr33}) with a \emph{flip}, realized with a operator  
$U(\gamma=0,\ \forall a)$, as in the Remark after previous Theorem 2.1.

\bigskip

\noindent{PROOF} As it is implicit in its definition (see Section 4),
Grover's operator is  a \emph{flip} followed by a reflection of the 
$|v>$-component with respect to the $|v>$-axis, where $|v>:=N^{-1/2}|1, 
\dots , 1>$.

\bigskip

\noindent{REMARK}\ \ Theorem (2.1) shows that Grover's unitary and
the generalized ones presented in this paper are analogue, and the 
realizability of the former implies the realizability of the latter.

\noindent{REMARK}\ \ If in (\ref{gr34}) the identity (\ref{gr36})
holds then, remembering (\ref{gr24}) and (\ref{gr25}), we can rewrite 
(\ref{gr34}) inthe form:
\begin{equation}
|u>=\left|-\sin{\theta\over 2},\ {{\cos{\theta\over 2}}\over\sqrt{N-1}},\
\dots,\ {{\cos{\theta\over 2}}\over\sqrt{N-1}}\right\rangle\label{gr37a}
\end{equation}
which, up to a phase, is the most general form of a vector in $R^N$
with $N-1$ components equal.
\bigskip

\noindent{REMARK}\ \ A matrix rapresentation  of the operator $U
(\beta_0,\varepsilon)$  in the basis  $(|i>)$, with $\varepsilon_2=\varepsilon_1\varepsilon_4\varepsilon_3$ is:
$$U(\beta_0,\varepsilon)=
\varepsilon_2\left( \begin{array}{ccccc}
1+\varepsilon_3\beta_0-1&\varepsilon_5{\sqrt{1-\beta_0^2}\over\sqrt{N-1}}& \varepsilon_5
{\sqrt{1-\beta_0^2}\over\sqrt{N-1}}&\dots&
\varepsilon_5{\sqrt{1-\beta_0^2}\over\sqrt{N-1}}\\
\varepsilon_5{\sqrt{1-\beta_0^2}\over\sqrt{N-1}}&1-{{1+\varepsilon_3\beta_0}\over{N-1}}&
-{{1+\varepsilon_3\beta_0}\over{N-1}}&\dots&-{{1+\varepsilon_3\beta_0}\over{N-1}}\\
\varepsilon_5{\sqrt{1-\beta_0^2}\over\sqrt{N-1}}&-{{1+\varepsilon_3\beta_0}\over{N-1}}&
1-{{1+\varepsilon_3\beta_0}\over{N-1}}&\dots&-{{1+\varepsilon_3\beta_0}\over{N-1}}\\
\vdots&\vdots&&\ddots&\\
\varepsilon_5{\sqrt{1-\beta_0^2}\over\sqrt{N-1}}&-{{1+\varepsilon_3\beta_0}\over{N-1}}&
-{{1+\varepsilon_3\beta_0}\over{N-1}}&\dots&1-{{1+\varepsilon_3\beta_0}\over{N-1}}\\
\end{array}\right)=$$

$$=\varepsilon_21-\varepsilon_2\left( \begin{array}{cccc}
1-\varepsilon_3\beta_0&-\varepsilon_5{\sqrt{1-\beta_0^2}\over\sqrt{N-1}}&\dots&
-\varepsilon_5{\sqrt{1-\beta_0^2}\over\sqrt{N-1}}\\
-\varepsilon_5{\sqrt{1-\beta_0^2}\over\sqrt{N-1}}&{{1+\varepsilon_3\beta_0}\over{N-1}}&\dots
&{{1+\varepsilon_3\beta_0}\over{N-1}}\\
\vdots&\vdots&\ddots&\vdots\\
-\varepsilon_5{\sqrt{1-\beta_0^2}\over\sqrt{N-1}}&{{1+\varepsilon_3\beta_0}\over{N-1}}&\dots
&{{1+\varepsilon_3\beta_0}\over{N-1}}\\
\end{array}\right)=$$
$$=\varepsilon_21-\varepsilon_2\left( \begin{array}{cccc}
1-cos\ \theta&-{{sin\ \theta}\over\sqrt{N-1}}&\dots&
-{{sin\ \theta}\over\sqrt{N-1}}\\
-{{sin\ \theta}\over\sqrt{N-1}}&{{1+cos\ \theta}\over{N-1}}&\dots&{{1+cos\
\theta}\over{N-1}}\\
\vdots&\vdots&\ddots&\vdots\\
-{{sin\ \theta}\over\sqrt{N-1}}&{{1+cos\ \theta}\over{N-1}}&\dots&{{1+cos\
\theta}\over{N-1}}\\
\end{array}\right)=$$
$$=\varepsilon_21-\varepsilon_22\left( \begin{array}{cccc}
sin^2\ {\theta\over 2}&-{{sin\ {\theta\over 2}cos{\theta\over 2}}\over
\sqrt{N-1}}&\dots&
-{{sin\ {\theta\over 2}cos{\theta\over 2}}\over\sqrt{N-1}}\\
-{{sin\ {\theta\over 2}cos{\theta\over 2}}\over\sqrt{N-1}}&{{cos^2\
 {\theta\over 2}}\over{N-1}}&\dots&{{cos{\theta\over 2}}\over{N-1}}\\
\vdots&\vdots&\ddots&\vdots\\
-{{sin\ \theta}\over\sqrt{N-1}}&{{cos\ {\theta\over 2}}\over{N-1}}&
\dots&{{cos^2\ {\theta\over 2}}\over{N-1}}\\
\end{array}\right)=$$
\begin{equation}
=\varepsilon_2\left(1-2|u><u|\right)\label{gr38}
\end{equation}
with $|u>$ given by (\ref{gr34}) or (\ref{gr37a}).

\bigskip\bigskip

\section{Optimal Choice of the Parameters} 

\bigskip\bigskip

In this section we study the following generalization of Grover's problem: 
given a fixed vector $|a>=\sum_ia_i|i>$, we look for a unitary operator  
$U=U(\beta_0, \varepsilon)$  of the form discussed in sections (1)
and (2), which increases the probability of the $0$-th component of $|a>$, 
i.e. such that:
\begin{equation}
|a_0|<|(U(\beta_0, \varepsilon)|a>)_0|:=\left| < 0|U(\beta_0, \varepsilon)|a>
 \right|\label{gr39}
\end{equation}

\noindent{ DEFINITION 1}\ \ A unitary operator $U(\beta_0, \varepsilon)$
of the form discussed in sections (1) and (2) is an \emph{optimal amplifier} 
for the $0$-th component of $|a>$ if condition (\ref{gr39}) is satisfied and:
\begin{equation}
|(U(\beta_0, \varepsilon)|a>)_0|\leq|(U(\overline\beta_0, \overline\varepsilon)|a>)_0|
\label{gr40}
\end{equation}
$\forall\beta_0\in \left[ 0, 1\right];\ \forall\varepsilon:=(\varepsilon_1,\ \dots,\ \varepsilon_5)
\in\{\pm 1\}^5$.
If moreover:
\begin{equation}
|(U(\overline\beta_0, \overline\varepsilon)|a>)_0|=1\label{gr41}
\end{equation}
then we speek of an \emph{absolute optimal amplifier}.

\noindent{ THEOREM 2}\ \ Given a unit vector of the form:
\begin{equation}
|a_G>:=a_0 |0>+b\sum_{i\not=0}|i>\label{gr42}
\end{equation}
with $a_0\not=0$ and
\begin{equation}
a_0^2+(N-1)b^2=1;\label{gr43}
\end{equation} 
an absolute optimal amplifier exists and is defined by: $\beta_0=\pm a_0$. 

\noindent{PROOF}
From equations (\ref{gr22}), (\ref{gr23}), (\ref{gr26}) and (\ref{gr27}) we 
have:
$$U|a_G>:=U\left(a_0 |0>+b\sum_{i\not=0}|i>\right)=$$
$$=\varepsilon_1\varepsilon_4\left[\beta_0a_0+\varepsilon_3\varepsilon_5\sqrt{(N-1)(1-\beta_0^2)}b\right]
|0>+$$
$$+\varepsilon_2\left[b+\varepsilon_5{\sqrt{1-\beta_0^2}\over \sqrt{N-1}}a_0-\left(1+\varepsilon_3
\beta_0\right)b\right]\sum_{i\not=0}|i>=$$
$$=\varepsilon_1\varepsilon_4\varepsilon_3\left( cos\ \theta a_0+\sqrt{N-1}sen\ \theta\ b\right)
|0>+\varepsilon_2\left({{sen\ \theta}\over \sqrt{N-1}}a_0-cos\ \theta\ b\right)
\sum_{i\not=0}|i>$$
The amplitude of $|0>$ is extremal if:
$${\partial\over {\partial\ \theta}}\left( cos\ \theta a_0+\sqrt{N-1}sen\
 \theta\  b\right) =-sen\ \theta\ a_0+\sqrt{N-1}cos\ \theta\ b=0$$
and this is satisfied by a $\overline\theta$ such that:
\begin{equation}
tg\ \overline\theta\ =\sqrt{N-1}{b\over{a_0}}\label{gr43a}
\end{equation}
that gives: 
$$\varepsilon_3\beta_0=cos\ \overline\theta={{\varepsilon_6}\over\sqrt{1+tg^2\ \overline\theta}}=
\varepsilon_6a_0$$
and
$$\varepsilon_5\sqrt{1-\beta_0^2}=sin\ \overline\theta={{\varepsilon_7tg\ \overline\theta}\over
\sqrt{1+tg^2\ \overline\theta}}=\varepsilon_7\sqrt{N-1}b$$
where $\varepsilon_6,\varepsilon_7\in\{\pm 1\}$ and we used (\ref{gr43}).
From (\ref{gr43a}) we have  $\varepsilon_6=\varepsilon_7$.

Therefore we obtain:
$$a_0\mapsto \varepsilon_1\varepsilon_4\varepsilon_3\varepsilon_6\left(sen^2\ \overline\theta+cos^2\
\overline\theta\right)= \varepsilon_1\varepsilon_4\varepsilon_3\varepsilon_6$$
and
$$b\mapsto \varepsilon_2\varepsilon_6\left( {{sen\ \overline\theta cos\ \overline\theta}\over
\sqrt{N-1}}-{{sen\ \overline\theta cos\ \overline\theta}\over\sqrt{N-1}}\right)=0$$
Thus the extremal amplitudes correspond to probability $1$ and this completes 
the proof.

\bigskip

\noindent{REMARK}\ \ The absolute optimality of the previous Theorem
(3.2) refers to the case when all the components $a_k (k\not=0)$ are equal.
 However, for a general vector, an optimal amplifier will not be absolutely
 optimal. This fact will be apparent from the following theorem.

\bigskip

\noindent{ THEOREM 3}\ \ An optimal amplifier for a generic
vector $a$ of the form:
\begin{equation}
|a>:=\sum_{j=0}^N a_j |j>\label{gr44}
\end{equation}
with:
\begin{equation}
\sum_ja_j^2=1\label{gr45}
\end{equation}
exists if $\sum_{k\not=0}a_k\not=0$ and it is given by an operator 
(\ref{gr32}) with the following choice: \ $tg\ \theta=\ {{\sum_{k\not=0}a_k}
\over{a_0\sqrt{N-1}}}$.\bigskip

\noindent{PROOF}
From equations (\ref{gr22}), (\ref{gr23}), (\ref{gr26}) and (\ref{gr27}) we 
have:
$$U|a>:=U\sum_{j=0}^N a_j |j>=
\varepsilon_1\varepsilon_4\left(\beta_0a_0+\varepsilon_3\varepsilon_5{\sqrt{1-\beta_0^2}\over\sqrt{N-1}}
 \sum_{k\not=0}a_k\right)|0>+$$
$$+\varepsilon_2\sum_{i\not=0}\left(a_i+\varepsilon_5{\sqrt{1-\beta_0^2}\over \sqrt{N-1}}
a_0-{{1+\varepsilon_3\beta_0}\over{N-1}}\sum_{k\not=0}a_k\right)|i>=$$
$$=\varepsilon_1\varepsilon_4\varepsilon_3\left( cos\ \theta a_0+{{sen\ \theta}\over\sqrt{N-1}}
\sum_{k\not=0}a_k\right) |0>+$$
$$+\varepsilon_2\sum_{i\not=0}\left(a_i+{{sen\ \theta}\over \sqrt{N-1}}a_0-{{1+
cos\ \theta}\over
{N-1}}\sum_{k\not=0}a_k\right)|i>=$$
and the amplitude of $|0>$ is extremal for:
$${\partial\over {\partial\ \theta}}\left( cos\ \theta a_0+{{sen\ \theta}
\over\sqrt{N-1}} \sum_{k\not=0}a_k\right) =
-sen\ \theta a_0+{{cos\ \theta}\over\sqrt{N-1}} \sum_{k\not=0}a_k=0$$
then for a $\overline\theta$ such that:
$$tg\ \overline\theta\ ={{\sum_{k\not=0}a_k}\over{a_0\sqrt{N-1}}}$$
this gives:
$$\varepsilon_3\beta_0=cos\ \overline\theta\ ={{\varepsilon_6}\over\sqrt{1+tg^2\ \overline\theta}}=
{{\varepsilon_6}\over\sqrt{1+{{(\sum_{k\not=0}a_k)^2}\over{a_0^2(N-1)}}}}=
{{{\varepsilon_6}a_0\sqrt{N-1}} \over\sqrt{a_0^2(N-1)+(\sum_{k\not=0}a_k)^2}}$$
$$\sqrt{N-1}\gamma_0=sen\ \overline\theta={{\varepsilon_7tg\ \overline\theta}\over\sqrt{1+tg^2\
\overline\theta}}
={{\varepsilon_7\sum_{k\not=0}a_k} \over\sqrt{a_0^2(N-1)+(\sum_{k\not=0}a_k)^2}}$$
with $\varepsilon_6$, $\varepsilon_7\in {\pm 1}$ and $\varepsilon_6\varepsilon_7=1$, i.e. $\varepsilon_6=\varepsilon_7$.
This gives:
$$a_0\mapsto {{\varepsilon_1\varepsilon_4\varepsilon_3\varepsilon_6\left[ a_0^2(N-1)+(\sum_{k\not=0}a_k)^2
\right]}
\over{\sqrt{N-1}\sqrt{a_0^2(N-1)+(\sum_{k\not=0}a_k)^2}}}=$$
\begin{equation}
={{\varepsilon_1\varepsilon_4\varepsilon_3\varepsilon_6\sqrt{a_0^2(N-1)+(\sum_{k\not=0}a_k)^2}}\over
\sqrt{N-1}}= \varepsilon_1\varepsilon_4\varepsilon_3\varepsilon_6\sqrt{a_0^2+ {{\left(\sum_{k\not=0}a_k
\right)^2}\over{N-1}}}\label{gr46}
\end{equation}

Finally:
$$a_i\mapsto \varepsilon_2\left[a_i+{{\varepsilon_6}\over\sqrt{N-1}}{{\sum_{k\not=0}a_k}
\over\sqrt{a_0^2(N-1)+(\sum_{k\not=0}a_k)^2}}a_0-{{1+\varepsilon_6{{a_0\sqrt{N-1}}
 \over\sqrt{a_0^2(N-1)+ (\sum_{k\not=0}a_k)^2}}}\over{N-1}}\sum_{k\not=0}a_k
\right]=$$
\begin{equation}
=\varepsilon_2\left( a_i- {{\sum_{k\not=0}a_k}\over{N-1}}\right)\label{gr47}
\end{equation}
and this completes the proof.\bigskip

\noindent{REMARK}\ \ Obviously (\ref{gr46}) and (\ref{gr47}) shall be
as in Theorem (3.2) if the vector $|a>$ is of the form (\ref{gr42}).

\noindent{REMARK}\ \ The action of the \emph{optimal amplifier} found
in Theorem (3.3) can be described in the following way:
``For every $|a>$:
\begin{description}
\item{1.} we subtract from every $a_i$, $i\not= 0$, the average of all the 
components different from the $0$-th one: 
$$a_i\mapsto a_i-{{\sum_{k\not=0}a_k}\over{N-1}}$$
\item{2.} Then for the $0$'s component we have of course:
$$a_0\mapsto \sqrt{1-\sum_{i\not=0}\left(a_i-{{\sum_{k\not=0}a_k}\over{N-1}}
\right)^2}=$$
$$=\sqrt{1-\sum_{i\not=0}a_i^2-(N-1){{\left(\sum_{k\not=0}a_k\right)^2}
\over{(N-1)^2}}+
2{{\left(\sum_{i\not=0}a_i\right)\left(\sum_{k\not=0}a_k\right)}\over{N-1}}}=$$
$$=\sqrt{a_0^2+ {{\left(\sum_{k\not=0}a_k\right)^2}\over{N-1}}}$$
(where in the last step we used (\ref{gr45})) as in the (\ref{gr46})". 
\end{description}

\bigskip
\bigskip

\section{Grover's algorithm} 
\bigskip\bigskip

Grover, in [1], considers the following problem (cf also [2]):

\noindent{ PROBLEM}: Given a (quantum) system with a state
space ${\mathcal H}$ of dimension $N=2^n$. Let $\{0,1\}^N=\{ S_0, S_2,
\dots, S_{N-1}\}=:S$ be the set of states  represented as $n$
q-bit string $\in {\mathcal H}$.  Let be given a function:
$$C: S \mapsto \{ 0, 1\} $$
with the following property: there is only one state, say $S_v$, such 
that $C(S_v)=1$, while $C(S)=0\quad\forall S\not=S_v$. Construct a quantum 
computer algorithm which is able to find the unknown $S_v$ state with high 
(say $>.5$) probability.

It is always possible to rename the states so that $\{ S_1, \dots, S_N\}=
\{0, \dots, N-1\}$ and $S_v=0$. In these notations let be given a vector:
$$|a>:=\sum_ia_i|i>$$

The first step in Grover's algorithm is to construct an operator Z
  that \emph{flips} the $0$-component. In our notations: 
$$Z:=1-2|0><0|$$

Grover then defines: 
$$|\tilde a>:=Z|a>=-a_0 |0>+\sum_{i\not=0}a_i|i>$$

and chooses the unit vector in formula (\ref{gr33}) as follows:
\begin{equation}
|v>:={1\over\sqrt{N}}\sum_k|k>={1\over\sqrt{N}}|1, \dots , 1>\label{gr47a}
\end{equation}

This gives:
$$<v|\tilde a>={1\over\sqrt{N}}\sum_k<k|\left(-a_0 |0>+\sum_{i\not=0}a_i|i>
\right)=
{1\over\sqrt{N}}\left(-a_0 +\sum_{k\not=0}a_k\right)$$

Then, denoting $P:=|v><v|$, Grover introduces the unitary operator
$D|\tilde a>:=-1+2P$, whose action on $|\tilde a>$ is
given by:
$$D|\tilde a>:=\left(-1+2P\right)|\tilde a>=-|\tilde a>+
2<v|\tilde a>\ 
|v>=-|\tilde a>+{2\over\sqrt{N}}\left(-a_0 +\sum_{k\not=0}a_k\right)\ |v>$$
$$=\left[\left(1-{2\over N}\right)a_0+{2\over N}\sum_{k\not=0}a_k\right]|0>+
\sum_{i\not=0}\left[-a_i+{2\over N}\left(-a_0 +\sum_{k\not=0}a_k\right)\right]\
 |i>$$

Then:
\begin{equation}
a_0\mapsto {{N-2}\over N}a_0+{2\over N}\sum_{k\not=0}a_k=\varepsilon_1\left[a_0+
\eta (a)\right]\label{ggr1}
\end{equation}

\begin{equation}
a_i\mapsto -a_i+{2\over N}\left(-a_0+\sum_{k\not=0}a_k\right)=\varepsilon_2
\left[a_i+c(a)\right]\label{ggr2}
\end{equation}

If $a_k=a_h\ \ \forall k$, $h\not=0$ (the Grover's agorithm case) then:
$$a_0\mapsto {{N-2}\over N}a_0+{{2(N-1)}\over N}a_i$$
$$a_i\mapsto \left[-1+{{2(N-1)}\over N}\right]a_i-{2\over N}a_0$$

We can arrive to the same result working only \emph{via} operator algebra. 
Grover's unitary is therefore:
$$U_G:=DZ=-(1-2|v><v|)(1-2|0><0|)=$$
$$=-1+2|v><v|+2|0><0|-{4\over\sqrt{N}})|v><0|$$
and it acts on $|a>$ in the following way:
$$(-1+2|v><v|+2|0><0|-{4\over\sqrt{N}})|v><0|)|a>=$$
$$=\left|a_0+{2\over N} \sum_{j=0}^{N-1}a_j-{4\over N}a_0,\ \left(a_i+
{2\over N}\sum_{j=0}^{N-1}a_j-{4\over N}a_0\right)_{i=1, \dots , N-1}
\right\rangle=$$
\begin{equation}
=\left|\left(1-{2\over N}\right)a_0+{2\over N} \sum_{k\not=0}a_k,\
\left({2\over N}a_0-a_i+{2\over N} \sum_{k\not=0}a_k\right)_{i=1, \dots , N-1}
\right\rangle\label{ggr3}
\end{equation}

Comparing (\ref{ggr3}) with (\ref{gr46})and (\ref{gr47}) we conclude that, 
within the class $U(\beta_0, \varepsilon)$, Grover's unitary is not optimal,
 not only for a generic initial vector $|a>$, but even for the initial vector
 $|a_G>$ used by Grover himself and given by  (\ref{gr43}).

\bigskip
\bigskip

\section{A One Step Solution for Grover's Problem} 

In the notations of section (4) let us write $|j>$ for the state $S_j$ so 
that, in particular, $|S_v>=|v>$, and suppose moreover that $j=0,1, \dots,
 N-1$.

In Grover's algorithm one uses the unitary operator:
$$V|j>=(-1)^{C(S_j)}|j>$$
which is a self adjoint involution, i.e. $V=V^*$ and $V^2=
1$

We will consider the unitary operator:

$$V|j>=\left\{\begin{array}{c}
|j>\quad if\  j\not= 0\  or\  C(S_j)=0\ \ \\
|0>\quad if\  C(S_j)=1\qquad\qquad\ \ \\
|j>\quad if\ j=0\ and\ C(S_v)=1
\end{array}
\right.$$
which is also an involution ($V^2=1$).

Notice that $V|j>=\sum_ku_{jk}|k>$ and $u_{jk}=\delta _{jk}$ if $j\not=
 v,0$; $u_{vo}=1$, $u_{vk}=0$ for $k\not=v$, $u_{0v}=1$, $u_{0k}=0$ for 
$k\not=v$. Thus $V$ is a local operator in the sense of Grover [1],
section (5).
The action of $V$ can be compactly described by:

$$V|j>=(1-\delta_{1,C(S_0)})\delta_{0,C(S_j)}|j>+\delta_{1,C(S_j)}|0>+
(1-\delta_{1,C(S_0)})\delta_{j,0}+$$
\begin{equation}+{1\over 2}\sum_{k=1}\left[1-(-1)^{C(S_k)}\right]|k>
\label{ggr5}
\end{equation}
from which it is clear that the physical action of V is realized by
parallel computation of the values of C, exactly as in Grover's algorithm.

\noindent{ THEOREM}\ \ In Theorem (3.2) let us choose the initial vector
$|a_G>$ in 
(\ref{gr43}) so that $a_0=b={1\over\sqrt{N}}$ (i.e. we choose Grover's initial
vector) and let us denote U the corresponding absolute uptimal unitary
operator. Define:
$$U_{OPT}:=V^*UV$$
then $U_{OPT}$ is an absolute optimal amplifier for the component
$|S_v>$ of $|a_G>$.

\noindent{PROOF}\ \ Clear from theorem (3.2) and the definition of
U.

\bigskip\bigskip
\section{APPENDIX A: Direct proof that U verifies the Isometricity
Condition} 

Let us now verify that the set of conditions (\ref{gr4}), \dots, (\ref{gr8}) 
are also sufficient conditions. To this goal we check if the isometricity 
condition (\ref{gr11}) is satisfied by the operator U, given by
(\ref{gr3}) if the parameters satisfy (\ref{gr4}), \dots, (\ref{gr8}). 
Then, replacing (\ref{gr22}) and (\ref{gr23}), which are equivalent to 
(\ref{gr4}), \dots, (\ref{gr8}), into (\ref{gr10}) and the (\ref{gr12}), 
we find:
$$\eta (a)^2=(-1+\varepsilon_4\beta_0)^2a_0^2+{{1-\beta_0^2}\over {N-1}}
\left(\sum_{k\not=0}a_k\right)^2+2\varepsilon_4\varepsilon_3\varepsilon_5a_0(-1+\varepsilon_4\beta_0)
{\sqrt{1-\beta_0^2}\over \sqrt{N-1}}\sum_{k\not=0}a_k$$

$$2a_0\eta (a)=2(-1+\varepsilon_4\beta_0)a_0^2+2\varepsilon_4\varepsilon_3\varepsilon_5a_0
{\sqrt{1-\beta_0^2}\over \sqrt{N-1}}\sum_{k\not=0}a_k$$

$$\eta (a)^2+2a_0\eta (a)=(-1+\beta_0^2)a_0^2+{{1-\beta_0^2}\over {N-1}}
\left(\sum_{k\not=0}a_k\right)^2+2\varepsilon_3\varepsilon_5a_0\beta_0{\sqrt{1-\beta_0^2}
\over \sqrt{N-1}}\sum_{k\not=0}a_k$$

$$(N-1)c(a)^2=(1-\beta_0^2)a_0^2+{{(1+\varepsilon_3\beta_0)^2}\over{(N-1)^2}}
\left(\sum_{k\not=0}a_k\right)^2 -2\varepsilon_5a_0{\sqrt{1-\beta_0^2}\over
\sqrt{N-1}}(1+\varepsilon_3\beta_0)\sum_{k\not=0}a_k$$

$$2c(a)\sum_{k\not=0}a_k=2\varepsilon_5a_0{\sqrt{1-\beta_0^2}\over \sqrt{N-1}}
\sum_{k\not=0}a_k-2{{1+\varepsilon_3\beta_0}\over{N-1}} \left(\sum_{k\not=0}a_k
\right)^2$$

Therefore the isometricity condition (\ref{gr11}) is equivalent to: 
$$\eta (a)^2+2a_0\eta (a)=(\beta_0^2-1)a_0^2+{{1-\beta_0^2}\over {N-1}}
\left(\sum_{k\not=0}a_k\right)^2+2\varepsilon_3\varepsilon_5a_0\beta_0{\sqrt{1-\beta_0^2}
\over \sqrt{N-1}}\sum_{k\not=0}a_k=$$
$$=-\gamma=(-1+\beta_0^2)a_0^2 -{{(1+\varepsilon_3\beta_0)^2}\over{(N-1)^2}}
\left(\sum_{k\not=0}a_k\right)^2+ 2\varepsilon_5a_0{\sqrt{1-\beta_0^2}\over
\sqrt{N-1}}(1+\varepsilon_3\beta_0)\sum_{k\not=0}a_k+$$
$$-2\varepsilon_5a_0{\sqrt{1-\beta_0^2}\over \sqrt{N-1}}\sum_{k\not=0}a_k +
2{{1+\varepsilon_3\beta_0}\over{N-1^2}} \left(\sum_{k\not=0}a_k\right)^2$$
or equivalently:
$$(\beta_0^2-1)a_0^2+{{1-\beta_0^2}\over {N-1}} \left(\sum_{k\not=0}
a_k\right)^2+2\varepsilon_3\varepsilon_5a_0\beta_0{\sqrt{1-\beta_0^2}\over \sqrt{N-1}}
\sum_{k\not=0}a_k=$$
$$=(-1+\beta_0^2)a_0^2 +{{2(1+\varepsilon_3\beta_0)-(1+\varepsilon_3\beta_0)^2}
\over{(N-1)^2}} \left(\sum_{k\not=0}a_k\right)^2+$$
$$+ 2\varepsilon_5a_0{\sqrt{1-\beta_0^2}\over\sqrt{N-1}}(1+\varepsilon_3\beta_0-1)
\sum_{k\not=0}a_k$$
which is an identity for any choice of $\varepsilon_3$, $\varepsilon_4$, $\varepsilon_5$ and
this ends the proof.

%

\section{Bibliography} 

\begin{description}
\item{[1]} Lov K. Grover: Quantum Mechanics helps in searching for a 
needle in a haystack, Phys. Rev. Lett. 79, 325-328
\item{[2]} Boyer M, Brassard G, Hoyer P and Tapp A: Tight bounds on 
quantum searching (preprint quant-ph/9605034)
\item{[3]} Luigi  Accardi, Ruben Sabbadini:
On the Ohya--Masuda quantum SAT algorithm
Volterra preprint 2000,
\item{[4]} M. Ohya and N. Masuda, {\it $NP$ problem in Quantum
Algorithm\/}, quant--ph/9809075.
\item{[5]} M. Ohya, {\it Mathematical Foundation of Quantum Computer\/},
Maruzen Publ. Company, 1998.
\item{[6]} A. Yu. Kitaev:
Quantum measurement and the abelian stabilizer problem,
quant-ph/9511026, 20-11 (1995)
\item{[7]} M. Ohya, N. Watanabe:
On Mathematical treatment of Fredkin-Toffoli-Milburn gate, 
Physica D, 120 (1998) 206-213
\item{[8]}  I.V. Volovich, Quantum Computers and Neural Networks, Invited 
talk at the
International Conference on Quantum Information held at Meijo University,
4-8 Nov. 1997, Proc. of the Conference.
\item{[9]} I.V. Volovich, Models of quantum computers and decoherence problem,
preprint Vito Volterra N358, 1999.
\item{[10]} I.V. Volovich, Mathematical Models of Quantum Computers and 
quantum decoherence problem, in Volume dedicated to V.A. Sadovnichij. Moscow 
State 
University, 1999, to be published.
\item{[11]} I.V. Volovich, Quantum Kolmogorov machine, Invited talk at the
International Conference on Quantum Information held at Meijo University,
1-6 March 1999, Proc. of the Conference.
\item{[12]} I.V. Volovich,  ``Atomic Quantum Computer'', quant-ph/9911062;
Volterra preprint N. 403, 1999, Universit\`a degli Studi di Roma ``Tor 
Vergata''.
\item{[13]} I.V. Volovich:
Models of quantum computers and decoherence problem.
Volterra preprint N. 358, 1999, Universit\`a degli Studi di Roma ``Tor 
Vergata''.
\item{[14]} M. Ohya, I.V. Volovich:
Quantum computing, NP--complete problems and chaotic dynamics.
Volterra preprint N. 426, 2000, Universit\`a degli Studi di Roma ``Tor
Vergata''.      
\item{[15]} Luigi  Accardi, Masanori Ohya, Ruben Sabbadini:
in preparation

\end{description}

\end{document}